\documentclass[conference]{IEEEtran}
\IEEEoverridecommandlockouts
\usepackage{multirow}
\usepackage{tabularx}
\usepackage{cite}
\usepackage{amsmath,amssymb,amsfonts}
\usepackage{algorithmic}
\usepackage{graphicx}
\usepackage{amsmath}
\usepackage{textcomp}
\usepackage{xcolor}
\usepackage{soul, color, xcolor}
\usepackage{hyperref}
\usepackage{float}
\usepackage{caption}

\begin{document}

\title{Learning Two-factor Representation for Magnetic Resonance Image Super-resolution\\

\thanks{${\dagger}$ Indicates equal contribution.}
\thanks{$^*$ Indicates corresponding author. Email: supengxiang@ncu.edu.cn.}
}

\author{\IEEEauthorblockN{Weifeng Wei$^{\dagger}$}
\IEEEauthorblockA{\textit{School of Information Engineering} \\
\textit{Nanchang University}\\
Nanchang, China \\
416100220248@email.ncu.edu.cn}
\and
\IEEEauthorblockN{Heng Chen$^{\dagger}$}
\IEEEauthorblockA{\textit{School of Mathematics and Computer Science} \\
\textit{Nanchang University}\\
Nanchang, China \\
409100220064@email.ncu.edu.cn}
\and
\IEEEauthorblockN{Pengxiang Su$^*$}
\IEEEauthorblockA{\textit{School of Software} \\
\textit{Nanchang University}\\
Nanchang, China \\
supengxiang@ncu.edu.cn}
}

\maketitle

\begin{abstract}
Magnetic Resonance Imaging (MRI) requires a trade-off between resolution, signal-to-noise ratio, and scan time, making high-resolution (HR) acquisition challenging. Therefore, super-resolution for MR image is a feasible solution. However, most existing methods face challenges in accurately learning a continuous volumetric representation from low-resolution image or require HR image for supervision. To solve these challenges, we propose a novel method for MR image super-resolution based on two-factor representation. Specifically, we factorize intensity signals into a linear combination of learnable basis and coefficient factors, enabling efficient continuous volumetric representation from low-resolution MR image. Besides, we introduce a coordinate-based encoding to capture structural relationships between sparse voxels, facilitating smooth completion in unobserved regions. Experiments on BraTS 2019 and MSSEG 2016 datasets demonstrate that our method achieves state-of-the-art performance, providing superior visual fidelity and robustness, particularly in large up-sampling scale MR image super-resolution.
\end{abstract}


\section{Introduction}

Magnetic Resonance Imaging (MRI) is crucial in clinical diagnosis and monitoring, offering non-invasive, detailed examinations of internal structures. Generally, obtaining the high-quality MR image is particularly important for medical diagnosis. However, during image acquisition, a trade-off must be made among resolution, signal-to-noise ratio, and scan time, posing a challenge in obtaining high-resolution (HR) MR image. Therefore, reconstructing a HR MR image from low-resolution (LR) scans is a feasible approach. Early studies on MR image super-resolution (SR) use optimization 
\cite{optimization1, optimization2, optimization3} and \cite{CNN1, CNN2, CNN3, CNN4, CNN5, CNN6} interpolation methods \cite{interpolation}. Subsequently,  apply convolutional neural networks to learn the transformation from low to high resolution which improves the performance of SR. However, these methods rely on HR image for supervision and are sensitive to distribution gaps between training and testing data, limiting their applicability.

Recently, Neural Radiance Field (NeRF) \cite{nerf1} and its subsequent works \cite{nerf2, nerf3, nerf4, nerf5, nerf6} have demonstrated the powerful spatial modeling capabilities of Implicit Neural Representations (INR). Subsequently, some MR image SR methods based on INR \cite{nerf-sr1, nerf-sr2, cunerf, MCSR} have addressed the aforementioned issues. INR assume that HR MR image are implicit differentiable functions of 3D image space coordinates, while 2D thick slice LR images are sparse discrete samplings of this function. Multi-layer Perceptrons (MLPs) are then used to learn the continuous volumetric representation from the limited LR samples. Therefore, INR allows for the use of downsampled image of arbitrary resolution as input and does not require HR image for supervision. Motivated by this, we primarily focuses on MR image SR methods based on INR.

\begin{figure*}
	\centering
	\includegraphics[width=18cm]{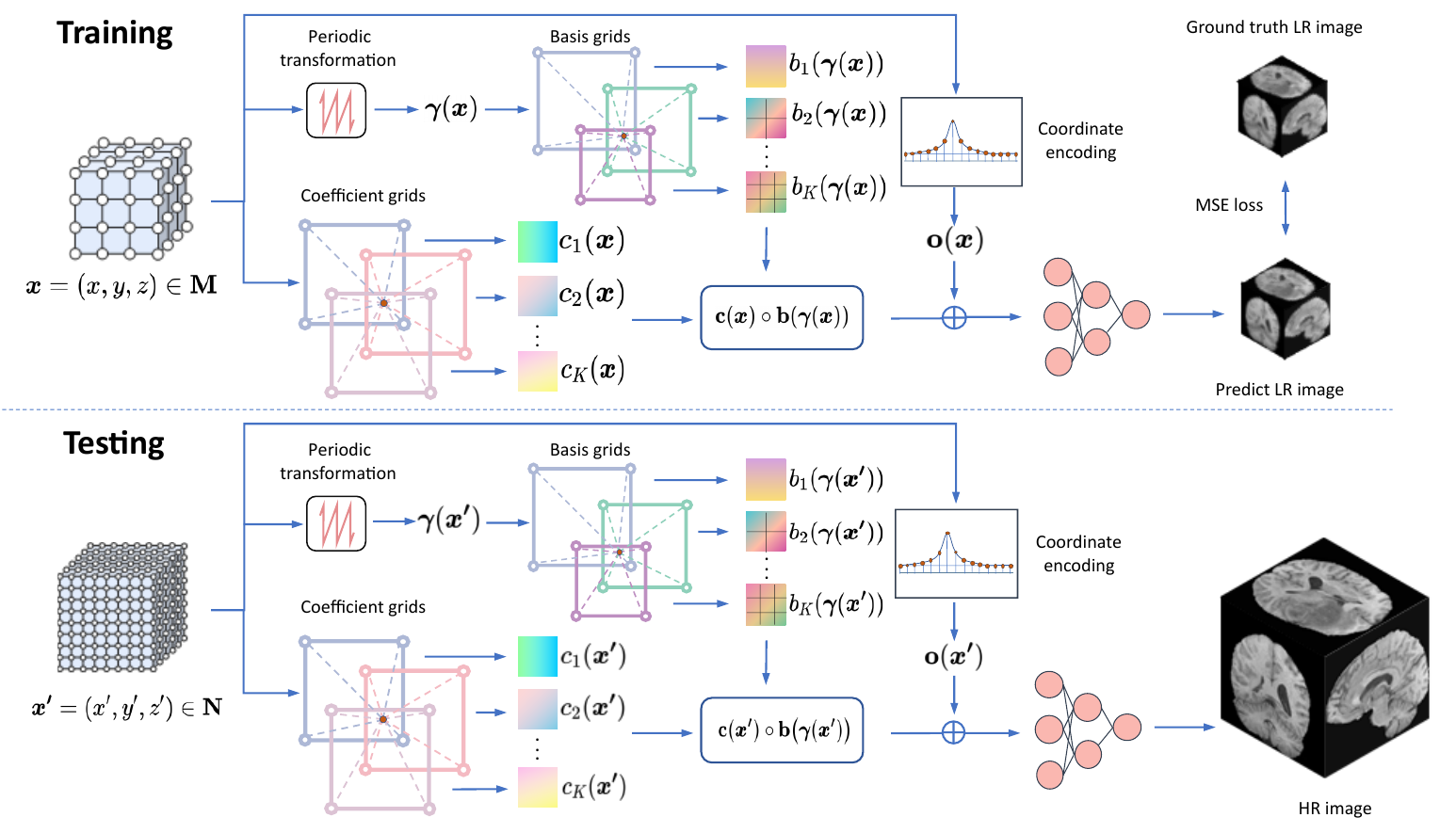}
	\caption{Overview of proposed method. The upper part shows the training process, and the lower part shows the testing process.} 
   \label{fig:1}
   \vspace{-0.35cm}
\end{figure*}

A major shortcoming of MR image SR methods \cite{nerf-sr1, nerf-sr2, cunerf, MCSR} based on INR is that they tend to apply Fourier Features \cite{fourier} to map input coordinates to higher-dimensional spaces for high-frequency signals capture. However, these approaches are relatively rigid and lacks flexibility, making it difficult to adapt to the multi-scale nature of signals. Since signals are often complex and structured, the fourier features remains a global method that struggles to efficiently handle features at different scales. In contrast, we propose employing the two-factor model to build the continuous volumetric representations for LR image. Specifically, we draw inspiration from \cite{dictionary1, dictionary2, dictionary3, SCARF, factorfield} and combine dictionary decomposition techniques with neural implicit representations. Different from \cite{nerf-sr2}, for the intensity signals in MR image, we decompose it into basis and coefficient factors. The basis factors promote structured signal features shared across scales, while the coefficient factors allow for spatially varying content. This novel neural representation allows efficient signals recovery from sparse observations, which achieves high-quality HR image reconstruction.

Another significant limitation of related work \cite{nerf-sr1, nerf-sr2, cunerf, MCSR} is that they rely on learning continuous volumetric representation from dense visual observations. However, LR MR image typically contain sparse visual information and more unobserved regions. Due to the lack of sufficient data support, such methods often struggle to generate reasonable predictions, leading to the appearance of holes. Although CuNeRF \cite{cunerf} attempt to alleviate this issue through super-sampling, the lack of effective supervisory signals in the samples makes it difficult for the model to learn continuous structures as the up-sampling scale increases. To address the problem of holes, we use a coordinate-based One-Blob \cite{one-blob} encoding to capture the structural relationship between sparse voxels accurately, which facilitates the smooth completion of unobserved regions without any super-sampling.

In summary, our contributions include: (1) We factorize the image intensity signals into learnable basis and coefficient factors, the linear combination of two-factor allows for efficient representation of both global and local attributes of the signal. (2) We perform coordinate encoding to the voxels of the input image, allowing the neural network to fully utilize the continuity and contextual information of the input data, which improves the flexibility and accuracy of the model in processing continuous input. (3) Experiments on two commonly used MRI datasets demonstrate that our method can achieve state-of-the-art performance in MR image SR.

\section{Method}
Given a rigidly registered 3D HR image, we consider down-sampling the image along three orthogonal scanning views (axial, coronal, sagittal) to obtain LR image, and then use the corresponding affine matrix to calculate the voxel coordinates set $\mathbf{M} = \{ (x, y, z) \mid (x, y, z) \in \mathbb{R}^3 \}$ in the scanner's reference space. Subsequently, $\mathbf{M}$ and the intensity of each voxel are normalized, where the voxel coordinates are input into the proposed model, which outputs the corresponding target intensity values. During the training stage, the two-factor representation is optimized by minimizing the mean squared error (MSE) loss between the target intensity values and the intensity values of the ground truth LR image. In the testing stage, we input the voxel coordinates of the coordinate set $\mathbf{N} = \{ (x', y', z') \mid (x', y', z') \in \mathbb{R}^3 \}$ of the HR image and fix all learnable parameters to evaluate the SR performance of proposed model. The overview of proposed method is shown in Fig.\ref{fig:1}.

\subsection{Two-Factor Representation}\label{AA}
Although previous methods \cite{nerf-sr1, cunerf, MCSR} introduce fourier features \cite{fourier} to address the implicit smoothing bias of MLPs for modeling high-frequency signals, this simplistic representation poses challenges in modeling multi-scale complex signals. In contrast, we explore the two-factor modeling of intensity signals. For a 3D voxel coordinate $\boldsymbol{x}  = (x, y, z) \in \mathbf{M}$, our goal is to learn a continuous volumetric representation through dictionary decomposition: \begin{equation}  
\mathbf{s}(\boldsymbol{x}) = \mathbf{c}(\boldsymbol{x})^\top \mathbf{b}(\boldsymbol{x})\label{eq:1}
\end{equation} where $\mathbf{c}(\boldsymbol{x})$ and $\mathbf{b}(\boldsymbol{x})$ represent the coefficient and basis factors of the 3D coordinate $\boldsymbol{x}$, respectively, while $\mathbf{s}(\boldsymbol{x})$ denotes the intensity signal at coordinate $\boldsymbol{x}$. The basis factors are responsible for capturing global and recurring features in the signal, achieving efficient signal representation by sharing basis across different positions and scales. Meanwhile, the coefficient factors focuses on the expression of local features, adapting to the local detail variations of the signal by flexibly adjusting the weights of the basis at different positions. For spatially varying coefficient factors, we achieve $\mathbf{c}(\boldsymbol{x})$ by performing trilinear interpolation in a total of $K$ levels tensor grid with constant resolution: \begin{equation}  
\mathbf{c}(\boldsymbol{x}) = \left({c}_1(\boldsymbol{x}), \ldots, {c}_K(\boldsymbol{x})\right)^\top\label{eq:2}\end{equation} \begin{equation}{c}_i(\boldsymbol{x}) = \text{TriLerp}(\boldsymbol{x}, \mathbf{G}_i) \quad i = 1,...,K\label{eq:3}\end{equation} where $\text{TriLerp}(.,.)$ indicates tri-linear interpolation and $\mathbf{G}_i$ is the i-th level of coefficient grids. We uniformly set the resolution of the coefficient grids to 32.

In addition, we implement the basis factors using multi-resolution tensor grids, allowing us to capture both fine details and smooth components of the signal across multiple scales. We also set up a total of $K$ levels of basis grids with each level increasing in resolution progressively: \begin{equation}  
\mathbf{b}(\boldsymbol{x}) = \left({b}_1(\boldsymbol{x}), \ldots {b}_K(\boldsymbol{x})\right)^\top\label{eq:4}\end{equation} the resolution of each level $R_i$ in basis grids is defined as follows: \begin{equation}
R_i = \left\lfloor R_{\text{min}} g^i \right\rfloor \quad g = \exp\left(\frac{\ln R_{\text{max}} - \ln R_{\text{min}}}{K - 1}\right)\label{eq:5}
\end{equation} where $R_{\text{min}}$ and $R_{\text{max}}$ represent the minimum and maximum resolutions, respectively. In this context, $R_{\text{min}} = 32$, $R_{\text{max}} = 128$, and there are $K = 6$ levels in total. Therefore, the i-th component of the basis factor can be expressed as: \begin{equation}b_i(\boldsymbol{x}) = \text{TriLerp}(\boldsymbol{x}, \mathbf{H}_{i} ) \quad i = 1,...,K\label{eq:6}\end{equation} where $\mathbf{H}_i$ is the i-th level of basis grids.

Before applying the voxel coordinates to the basis grids, we need to implement a periodic transformation $\boldsymbol{\gamma}(\boldsymbol{x})$ on them. The periodic transformation supports the continuity and consistency of the basis factors across various positions and scales of the signal. This consistency is crucial for multi-scale volumetric representation, as it ensures that the signal is uniformly processed and represented at different scales. In this paper, the periodic transformation function $\boldsymbol{\gamma}$ is defined as follows: \begin{equation}
\boldsymbol{\gamma}(\boldsymbol{x}) = \left( \text{Sawtooth}(\boldsymbol{x}, 1), \dots, \text{Sawtooth}(\boldsymbol{x}, K) \right) \label{eq:7}\end{equation} \begin{equation}
\text{Sawtooth}(\boldsymbol{x}, i) = (\boldsymbol{x} \cdot f_i) \mod \left( 2 / f_i \right) \quad i = 1,...,K \label{eq:8}\end{equation} where $\text{Sawtooth}(.,.)$ is a sawtooth function with a variable period, $f_i$ represents the frequency at level $i$, and we define $f_i = 2+1.2(i-1)$. This frequency is used to control the coverage of the basis factors in the target signal region. Then, Eq.\ref{eq:1} and Eq.\ref{eq:4} can be expanded to: \begin{equation}\mathbf{s}(\boldsymbol{x}) = \mathbf{c}(\boldsymbol{x}) \circ \mathbf{b}\left(\boldsymbol{\gamma}(\boldsymbol{x})\right)\label{eq:9}
\end{equation} \begin{equation}  
\mathbf{b}(\boldsymbol{\gamma}(\boldsymbol{x})) = \left({b}_1(\boldsymbol{\gamma}(\boldsymbol{x})), \ldots {b}_K(\boldsymbol{\gamma}(\boldsymbol{x}))\right)^\top \quad i = 1,...,K\label{eq:10}\end{equation} where $\circ$ indicates the element-wise product, which helps to address the limited band-width of MLPs.

\subsection{Coordinate-based Encoding}
In the SR task of MR image, the higher the voxel sparsity of the LR image tend to difficult to reconstruct. INR tend to learn a continuous volume representation in dense inputs, but have limited generalization in overly sparse samples, which easily leaves holes in unobserved areas when restoring HR image. Especially when the up-sampling scale increases, this approach has difficulty capturing the structural relationship between voxels. We find that smoothing the unobserved area is an effective solution. To this end, we apply One-Blob \cite{one-blob} encoding to input voxel coordinates. Specifically, we generate a feature with a strong local response and a rapid decay away from the coordinate for each input sparse voxel coordinate $\boldsymbol{x}$: \begin{equation}
\mathbf{o}(\boldsymbol{x}) = \text{One-Blob}(\boldsymbol{x})\end{equation} where $\mathbf{o}(\boldsymbol{x})$ is the feature generated by the One-Blob encoder. This enables the model to learn detailed information about the intensity signal based on these local responses near a certain coordinate. The decay of these local responses ensures that the influence of a point is primarily concentrated in its local neighborhood, without extending to more distant regions 
\begin{figure*}{}
	\centering
 \rotatebox{90}{\scriptsize{~~~~~ MSSEG 2016 \cite{MSSEG}~~~~~~~~BraTS 2019 \cite{BraTS}}}
	\includegraphics[width=17cm]{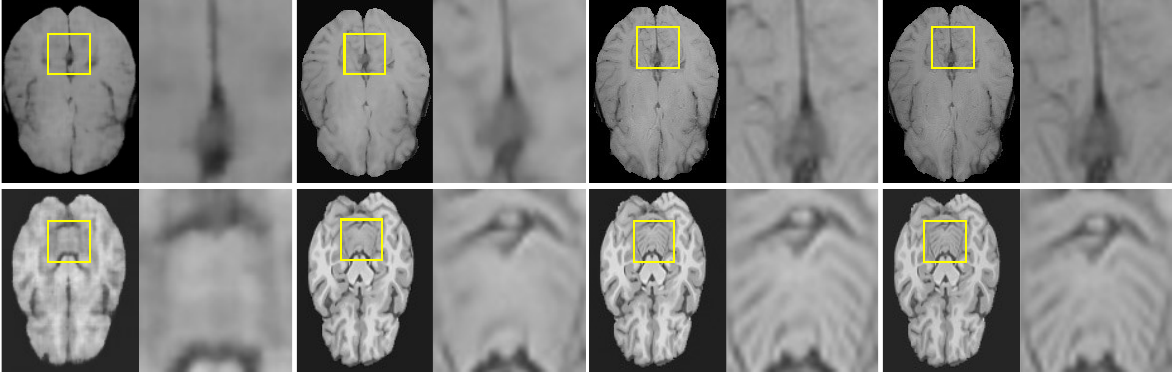}
  \rotatebox{0}{\scriptsize{~ CuNeRF \cite{cunerf}~~~~~~~~~~~~~~~~~~~~~~~~~~~~~~~~~MCSR \cite{MCSR}~~~~~~~~~~~~~~~~~~~~~~~~~~~~~~~~~~~~~~~~Ours~~~~~~~~~~~~~~~~~~~~~~~~~~~~~~~~~~~ Ground Truth}}
	\caption{Comparison of qualitative results on BraTS 2019 \cite{BraTS} and MSSEG 2016 \cite{MSSEG} datasets. We set up-sampling scale to 4.} 
   \label{fig:2}
\end{figure*}

\begin{table*}[htb]
\footnotesize
\centering  
\captionsetup{labelsep=space}
\caption{Quantitative comparison on BraTS 2019 \cite{BraTS} and MSSEG \cite{MSSEG} datasets.}  
\label{tab:1}  
\begin{tabularx}{\textwidth}{|c|*{8}{>{\centering\arraybackslash}X|}} 
\hline  
\multirow{4}{*}{Method}  
                        & \multicolumn{4}{c|}{BraTS \cite{BraTS}}        & \multicolumn{4}{c|}{MSSEG 2016 \cite{MSSEG}} \\ \cline{2-9}  
                        & \multicolumn{2}{c|}{$\sigma = 4$}      & \multicolumn{2}{c|}{$\sigma = 8$} & \multicolumn{2}{c|}{$\sigma = 4$} & \multicolumn{2}{c|}{$\sigma = 8$} \\ \cline{2-9}  
                        & PSNR (↑)  & SSIM(↑) & PSNR(↑) & SSIM(↑) & PSNR(↑) & SSIM(↑) & PSNR(↑) & SSIM(↑) \\ \hline  
Tricubic \cite{interpolation}           & 21.29      & 0.907    & 18.65    & 0.849    & 21.59    & 0.934    & 18.24    & 0.817    \\ \hline  
LRTV \cite{optimization3}                   & 21.46      & 0.912    & 19.31    & 0.867    & 22.64    & 0.947    & 18.87    & 0.826    \\ \hline  
CuNeRF \cite{cunerf}                 & 22.54      & 0.921    & 20.49    & 0.916    & 21.53    & 0.935    & 19.32    & 0.849    \\ \hline  
MCSR \cite{MCSR}                   & 28.89      & 0.972    & 25.85    & 0.938    & 24.91    & 0.959    & 21.30    & 0.916    \\ \hline
Ours                    & \textbf{32.57}     & \textbf{0.976}   & \textbf{30.98}    & \textbf{0.954}    & \textbf{30.02}    & \textbf{0.968}    & \textbf{27.93}    & \textbf{0.951}    \\\hline   
\end{tabularx}  
\end{table*} 
or interfering with other voxels, which can avoid the global blurring caused by over-smoothing.

Therefore, when the proposed model learns on sparse voexl coordinates, we can generate smooth transitions between these coordinates through its local properties. This local sensitivity allows the model to learn a continuous volumetric representations even in LR image with sparse data. It not only helps in generating high-quality representations near the observed data but also enables filling in gaps in unobserved regions through smooth transitions, thereby ensuring the coherence and accuracy of the volumetric representation.

\subsection{Model Optimization}
Based on the above analysis, we use a 2-layer tiny-MLP of size 64 to regress the intensity signal $s(\boldsymbol{x})$ and feature $o(\boldsymbol{x})$ of voxel coordinate $\boldsymbol{x}$ to the target intensity value: \begin{equation} \hat{I}(\boldsymbol{x}) = \text{MLP}(s(\boldsymbol{x}), \mathbf{o}(\boldsymbol{x})) \end{equation} then, we optimize our model by minimizing the MSE loss function between the target voxel intensity $\hat{I}$ and the ground truth voxel intensity $I$ of the LR image. The loss function $L$ is expressed as: \begin{equation}\mathcal{L}(\theta) = \sum_{\boldsymbol{x} \in M} \left( \hat{I}(\boldsymbol{x}) - I(\boldsymbol{x}) \right)^2
\end{equation}
where $\theta$ is all learnable parameters.

\section{Experiments}
\textbf{Datasets}. We conduct experiments on two commonly used MRI datasets: BraTS 2019 \cite{BraTS} and MSSEG 2016 \cite{MSSEG}, then we select HR scans (T1-weighted) of 50 patients from each of these two datasets. For a given up-sampling scale $\sigma$, the LR image is obtained by down-sampling from $H \times W \times D$ (HR) to $\frac{H}{\left\lfloor \sqrt[3]{\sigma} \right\rfloor} \times \frac{W}{\left\lfloor \sqrt[3]{\sigma} \right\rfloor} \times \frac{D}{\left\lfloor \sqrt[3]{\sigma} \right\rfloor}$ (LR), which ensures that the voxels of the LR image are not densely distributed along a certain view.

\textbf{Baseline}. The baselines used for comparison including traditional Tricubic interpolation method \cite{interpolation}, Optimization-based methods LRTV \cite{optimization3} and INR based methods CuNeRF \cite{cunerf}, MCSR\cite{MCSR}.

\textbf{Metrics}. We use two commonly used quantitative metrics in MR image SR tasks, namely Peak Signal-to-Noise Ratio (PSNR) and Structured Similarity Index Measure (SSIM) \cite{ssim}, to assess the image quality of different methods. In our experiments, we conduct quantitative evaluations on completed 3D MRI volume rather than 2D slices within it.

\textbf{Setting}. Our method employs the Adam optimizer \cite{adam} with a batch size of 1000 and 50 epochs, consistent with \cite{cunerf, MCSR}. The initial learning rates for the tensor grids and MLP are set to 2e-2 and 1e-3, respectively, with a weight decay of 5e-4.

\subsection{Experimental Results}
We have detailed the results obtained using two up-sampling scales on two 3D MRI datasets in Tab.\ref{tab:1}, and all quantitative results are the average of the evaluation metrics of 50 patients in the two datasets. As shown in the Tab.\ref{tab:1}, our method essentially outperforms the baseline methods for MR image SR tasks. In addition, we visualize the MR image SR results of the proposed method and other baseline methods on the BraTS 2019 \cite{BraTS} and MSSEG 2016 \cite{MSSEG} datasets in Fig.\ref{fig:2}. Compared to the baseline methods, the proposed method more accurately recovers detailed textures and complex structures in HR image, achieving better visual fidelity and reducing holes and blur.

\subsection{Ablation Study}
We demonstrate our design choices through ablation experiments on two up-sampling scales on the BraTS 2019 dataset \cite{BraTS}, and report quantitative results for removing individual components and the complete model in Tab.\ref{tab:2}. The results show that using only one factor and removing the periodic transformation yields sub-optimal results. At the same time, removing coordinate-based one-blob encoding significantly impacts the results on larger up-sampling scales. This further demonstrates that the joint two-factor representation can achieve higher-quality reconstruction of HR image, while the coordinate-based encoding enhances the robustness of SR task.
\begin{table}[htbp]
\centering
\captionsetup{labelsep=space}
\caption{Ablation results on BraTS 2019 Dataset \cite{BraTS}.} 
\begin{tabular}{|c|cc|cc|}
\hline
\multirow{2}{*}{Method} & \multicolumn{2}{c|}{$\sigma = 4$}             & \multicolumn{2}{c|}{$\sigma = 8$}             \\ \cline{2-5} 
                        & \multicolumn{1}{c|}{PSNR(↑)} & SSIM(↑) & \multicolumn{1}{c|}{PSNR(↑)} & SSIM(↑) \\ \hline
w/o basis factors        & \multicolumn{1}{c|}{24.42}    & 0.931    & \multicolumn{1}{c|}{21.91}    & 0.892    \\ \hline
w/o coeff. factors        & \multicolumn{1}{c|}{30.38}    & 0.943    & \multicolumn{1}{c|}{28.81}    & 0.923    \\ \hline
w/o per. trans.         & \multicolumn{1}{c|}{31.03}    & 0.945    & \multicolumn{1}{c|}{29.57}    & 0.925    \\ \hline
w/o one-blob enc.        & \multicolumn{1}{c|}{27.29}    & 0.914    & \multicolumn{1}{c|}{23.69}    & 0.909    \\ \hline
complete model          & \multicolumn{1}{c|}{\textbf{32.57}}    & \textbf{0.976}    & \multicolumn{1}{c|}{\textbf{30.98}}    & \textbf{0.954}    \\ \hline
\end{tabular}
\label{tab:2}
\vspace{-0.30cm}
\end{table}

\section{Conclusion}
In this paper, we propose a novel two-factor based representation for MR image super-resolution. We factorize the intensity signals into a linear combination of basis and coefficient factors to learn a continuous volumetric representation. In addition, we introduce a coordinate-based encoding in the implicit neural representation, making it more robust on large-scale super-resolution tasks. Experiments on two commonly used MRI datasets show that our method achieves state-of-the-art performance in neural implicit-based MR image super-resolution.
\section*{Acknowledgment}
This work was supported by National Natural Science Foundation of China, grant number~62262040.

\bibliographystyle{IEEEtran}
\bibliography{icassptemplate.bib}
\vspace{12pt}

\end{document}